# Teleparallelism as anholonomic geometry

D. H. Delphenich
Spring Valley, OH 45370
e-mail: feedback@neo-classical-physics.info

**Abstract.** – The geometry of parallelizable manifolds is presented from the standpoint of regarding it as conventional (e.g., Euclidian or Minkowskian) geometry, when it is described with respect to an anholonomic frame field that is defined on the space in question.

**1. Introduction.** – When one considers the era (1905-16) in which Einstein first developed his theory of relativity, and the somewhat-heuristic line of reasoning that led him to its final form [**1**], one must always ask the deeper question of whether that final form was the only one possible or simply the best-possible model in the context of that era. In particular, the dominant approach to the geometry of curved spaces at the time was Riemannian geometry, which was communicated to Einstein by his friend Marcel Grossmann. In a nutshell, Riemannian geometry is based in a metric tensor field and a metric connection with vanishing torsion that the metric defines.

In the decades that followed, many alternatives to Riemannian geometry were explored, all with their own successes and failures. For instance, Hermann Weyl considered extending beyond metric connections into the realm of conformal Lorentzian geometry [**2**], and Élie Cartan developed the theory of spaces with non-vanishing torsion [**3**], which later formed the basis for the work of Dennis Sciama [**4**] and Thomas Kibble [**5**] on extending the Einstein equations to space-times with torsion, which spawned a host of subsequent research on that topic.

There were also attempts to extend the scope of general relativity (i.e., gravity as a manifestation of the geometry of space-time) to higher-dimensional manifolds in the hopes of unifying Einstein's equations for gravitation with Maxwell's equations of electromagnetism, such as the attempts of Theodor Kaluza [**6**] and Oskar Klein [**7**]. Einstein and Schrödinger also considered generalizing from a symmetric covariant second-rank tensor field (viz., the Lorentzian metric) to an asymmetric one. (See the discussion of that theory in the book by Lichnerowicz [**8**], as well as the Kaluza-Klein approach, which he referred to as the "Jordan-Thiry" theory). There were also many attempts to embed Einstein's theory into the world of projective differential geometry [**9**-**11**].

One of the more intriguing avenues that was pursued in the name of the Einstein-Maxwell unification program was what Einstein and Mayer (and others) were calling "teleparallelism" ([1]). Eventually, the geometry of parallelizable manifolds, which is what teleparallelism amounts to mathematically, was shown to be equivalent to general relativity [**13**], in the sense that the field equations in either case admitted the same solutions. Indeed, one of the more intriguing aspects of the teleparallelism approach to gravitation is that it represents an extension of the gravito-electromagnetic approach to weak-field gravitation to strong-field gravitation [**14**].

---

([1]) The author has compiled an anthology [**12**] of English translations of many of the early papers by Einstein, Mayer, Cartan, Bortolotti, Zaycoff, and Stiefel on this topic. It is currently available as a free PDF download at the author's website: neo-classical-physics.info.

What makes teleparallelism somewhat more fundamental in character than other, more heuristic, approaches to space-time geometry is that the question of the parallelizability of a differentiable manifold is quite topologically deep, so in effect, one must always address it eventually as a topological issue. However, it is also fundamental geometrically, because the whole purpose of introducing connections on manifolds (or rather their bundles of tangent vectors or tangent frames) is to have some way of defining the parallelism of finitely-spaced tangent objects. Typically, one can only hope for an infinitesimal association that becomes finite by integration, and generally only along connecting paths.

However, any point of a differentiable manifold admits some neighborhood on which one can define a local frame field, and if one *defines* that local frame field to be "parallel" then one can speak of the parallelism of tangent objects at any pairs of finitely-space points that lie within that neighborhood. Now, one of the drawbacks to the usual way that "local" objects are introduced on manifolds is that "local" is always treated in the analytical sense, as if it always mean "in a sufficiently-small neighborhood of the point," whereas the maximum domain of definition of a local frame field might very well be the complement of a single point. For instance, that is the case on a two-sphere, which admits a tangent 2-frame field everywhere except for one point. In such a case, one might consider that point to be a geometric singularity, as well as a topological one, which points the way to "singular teleparallelism" (cf., the author's comments on this in the anthology [**12**]). Furthermore, there are differentiable manifolds, such as Lie groups, that admit global frame fields without admitting global coordinate systems, so vector spaces and affine spaces are not the only examples of parallelizable manifolds. For instance, any compact Lie group is parallelizable, but not diffeomorphic to $\mathbb{R}^n$, which is not compact.

Hence, one might consider the geometry of parallelizable manifolds in its own right, if only to get some sense of how it relates to fundamental, if not elementary, issues in geometry. That is essentially the purpose of the present study, namely, to treat teleparallelism (viz., the geometry of parallelizable manifolds) as an approach to geometry that is rooted in elementary geometric concepts, and not just a way of transforming Einstein's equations.

Although the discussion will start out at the level of basic linear algebra, nonetheless, past that point, a certain familiarity with the modern geometric methods in mathematical physics will be assumed, at least the basic definitions of differentiable manifolds and Lie groups and the methods of tensor analysis; in particular, the calculus of exterior differential forms ([1]), which are closely related to Cartan's approach to differential geometry in terms of moving frames [**17**].

After the linear-algebraic preliminaries of § **2**, which are mostly intended to define the concepts of frames and coframes in vector spaces and their duals, the geometry of parallelizable spaces will be presented in § **3** as a branch of elementary geometry in which the existence of a global frame field on a manifold can be used to define not just the parallelism of tangent and cotangent geometric objects, but also a volume element and a metric tensor of any desired signature type. § **4** is devoted to developing the various canonical constructions that one can derive from an anholonomic global frame field from an analytical and algebraic standpoint. In § **5**, the usual approach to teleparallelism will be discussed, in which one introduces the connection (on the frame bundle) that makes the global frame field parallel in the usual differential-geometric sense. It is then shown that the properties of that connection are consistent with the elementary facts that were presented in § **3**, and more to the point, with the canonical constructions that always pertain to anholonomic frame fields. Finally, in § **6**, some of the directions for further development are discussed.

---

  ([1]) A good treatment of these topics that is oriented towards physicists is the book by Theodore Frenkel [**15**]. For a discussion of the geometry of parallelizable manifolds, one can confer Bishop and Crittenden [**16**], in which the discussion takes the form of a series of problems for the reader.

Ultimately, the conclusion of this study will be that the geometry of parallelizable manifolds is essentially elementary geometry when it is described with respect to an anholonomic frame field. Note that there is then a fundamental difference between using a frame field as one's reference frame for space and using a coordinate system, which essentially implies the integrability of the frame field, which is not always possible.

**2. Linear-algebraic preliminaries.** – We shall start by reviewing the linear-algebraic notions that are at the foundations of parallelizations of spaces.

*a. Frames and coframes.* – A *frame* in an $n$-dimensional vector space $V$ is a basis for $V$, namely, a set $\{\mathbf{e}_i, i = 1, \ldots, n\}$ of $n$ linearly-independent vectors in $V$. Hence, any vector $\mathbf{v} \in V$ can be expressed uniquely in terms of its components $v^i$ with respect to that frame ([1]):

$$\mathbf{v} = \mathbf{e}_i v^i. \tag{2.1}$$

If $V^*$ is the dual space to $V$ (i.e., the space of linear functionals – or covectors – on $V$) then a frame in $V^*$ is a set $\{\theta^i, i, \ldots, n\}$ of $n$ linearly-independent covectors, and is usually referred to as a *coframe*, for that reason. Hence, any covector $\alpha \in V^*$ can be expressed uniquely by its components $\alpha_i$ with respect to a chosen coframe:

$$\alpha = \alpha_i \theta^i. \tag{2.2}$$

Any frame $\mathbf{e}_i$ on $V$ is associated with a unique coframe $\theta^i$ on $V^*$ that is called its *reciprocal coframe*, and is defined by the requirement:

$$\theta^i(\mathbf{e}_j) = \delta^i_j. \tag{2.3}$$

That means that a choice of frame on $V$ will always define a linear isomorphism of $V$ and $V^*$ that takes $\mathbf{v}$ as in (2.1) to the covector $v^i \theta^i$, although one can see by the position of the indices that this isomorphism is only so useful if one wishes that the components should transform contragrediently to the coframe.

---

 ([1]) Since we will be dealing with both right and left actions of matrices on frames and coframes, it will be convenient for us to put components of vectors and covectors on the side that the frame changes act upon. That will have the effect of always making the summations over identical indices behave like matrix multiplications.

If a vector **v** is expressed as in (2.1) and $\theta^i$ is the reciprocal coframe to the frame $\mathbf{e}_i$ then one will see that:

$$\theta^i(\mathbf{v}) = \theta^i(\mathbf{e}_j\, v^j) = \theta^i(\mathbf{e}_j)\, v^j = \delta^i_j\, v^j = v^i. \tag{2.4}$$

In other words, the action of the coframe elements is to pick off the components of any vector with respect to its reciprocal frame.

If $\bar{\mathbf{e}}_i$ is another frame on $V$ then since each vector of $\bar{\mathbf{e}}_i$ can be expressed in terms of its components $\tilde{h}_i^j$ with respect to an initial choice of frame, such as $\mathbf{e}_i$, one can express the transition from $\mathbf{e}_i$ to $\bar{\mathbf{e}}_i$ as the system of $n$ linear equations in $n$ unknowns:

$$\bar{\mathbf{e}}_i = \mathbf{e}_j\, \tilde{h}_i^j. \tag{2.5}$$

Since both sets of $n$ vectors are linearly-independent, the $n \times n$ matrix $\tilde{h}_i^j$ is invertible, and therefore belongs to the group $GL(n, \mathbb{R})$.

Dually, if $\theta^i$ and $\bar{\theta}^i$ are two coframes on $V^*$ then one can express the one in terms of the other by:

$$\bar{\theta}^i = h^i_j\, \theta^j, \tag{2.6}$$

in which the $n \times n$ matrix $h^i_j$ is also invertible. Furthermore, if $\bar{\mathbf{e}}_i$, as in (2.5), is reciprocal to $\bar{\theta}^i$ then:

$$\delta^i_j = \bar{\theta}^i(\bar{\mathbf{e}}_j) = (h^i_k\, \theta^k)(\mathbf{e}_l\, \tilde{h}_j^l) = (h^i_k\, \tilde{h}_j^l)\, \theta^k(\mathbf{e}_l) = (h^i_k\, \tilde{h}_j^l)\, \delta^k_l = h^i_k\, \tilde{h}_j^k.$$

Hence, $\tilde{h}_j^i$ must be the inverse of the matrix $h_i^j$, and we shall use the tilde from now on to indicate the inverse of a matrix. One then says that the reciprocal frames transform *contragrediently* under a change of frame.

Similarly, the components of a vector **v** and a covector $\alpha$ will transform contragrediently, but with the inverse matrices to the ones that change the frame and coframe:

$$\bar{v}^i = h^i_j\, v^j, \qquad \bar{\alpha}_i = \alpha_j\, \tilde{h}_i^j. \tag{2.7}$$

That is unavoidable if one wishes that the vector **v** and covector $\alpha$ should be defined independently of any choice of frame or coframe, as opposed to their components.

**3. Parallelizable manifolds.** – We next extend from $n$-dimensional vector spaces to $n$-dimensional differentiable manifolds, which we will assume to be smooth, so as to avoid having to deal with the analytical details that arise from the fact that differentiating something that is not smooth will lower its degree of differentiability.

*a. Frame and coframe fields.* – If an *n*-dimensional manifold *M* is differentiable then one can speak of the tangent space $T_xM$ at any point $x \in M$, which will be an *n*-dimensional vector space. Similarly, one can define its dual cotangent space $T_x^*M$, which is then an *n*-dimensional vector space of covectors at *x*. Hence, the linear-algebraic considerations above can be applied to the tangent and cotangent spaces to any differentiable manifold.

A (global) *frame field* on *M* is the smooth assignment of a frame $\mathbf{e}_i(x)$ in $T_x$ ([1]) with each $x \in M$. Dually, one can define a *coframe field* on *M* by a smooth assignment of a coframe $\theta_x^i$ in $T_x^*$ with each *x*. If $\pi : T(M) \to M$ and $\pi^* : T^*(M) \to M$ are the canonical projections of the tangent and cotangent bundles onto *M* then a frame field is a set of *n* linearly-independent sections $\mathbf{e}_i : M \to T(M)$, $x \mapsto \mathbf{e}_i(x)$ of the projection $\pi$ (i.e., *n* linearly-independent vector fields), while a coframe field is a set of *n* linearly-independent sections $\theta^i : M \to T^*(M)$, $x \mapsto \theta_x^i$ of the projection $\pi^*$ (i.e., *n* linearly-independent covector fields). Hence, by the definition of a section:

$$\pi \cdot \mathbf{e}_i = I, \qquad \pi^* \cdot \theta^i = I, \qquad \text{for all } i,$$

in which *I* is the identity map on *M*. In effect, all that this says is that $\mathbf{e}_i(x)$ will be an element of $T_x$ for each *i* and $\theta_x^i$ will be an element of $T_x^*$.

If one denotes the differentiable manifold of all tangent linear frames at *x* by $GL_x(M)$, or just $GL_x$, and the bundle of linear frames on *M* by $GL(M) \to M$ (so its fiber at *x* will be $GL_x$) then one can regard a frame field on *M* as a smooth section $\mathbf{e} : M \to GL(M)$, $x \mapsto \mathbf{e}_i(x)$ of the bundle projection. Dually, if $GL_x^*$ is the differentiable manifold of cotangent linear coframes at *x* and $GL^*(M) \to M$ is the corresponding bundle projection then a coframe field will be a smooth section $\theta : M \to GL^*(M)$, $x \mapsto \theta_x^i$ of the latter projection.

If a differentiable manifold *M* has a global frame field $\mathbf{e}_i(x)$ (or dually, a global coframe field $\theta_x^i$) defined on it then *M* is said to be *parallelizable*. Other terminologies that one encounters in the literature are that it admits *distant parallelism, absolute parallelism*, and that it is a *Weitzenböck space*.

One of the growing tendencies of modern differential geometry is to always revert to differential topology, which is probably due to the possibility that the usual definition of a differentiable manifold is too general to be a useful basis for any geometry that is not purely general in scope, so one always has to discuss the topological obstructions to the parallelizability of a manifold, which go back to the thesis [18] of Eduard Stiefel under Heinz Hopf, and can be summarized by saying that a necessary (but not sufficient) condition for the parallelizability of a manifold is that its Stiefel-Whitney classes should all vanish [19]. We do not wish to get sidetracked with topology here, so we shall simply give some examples of manifolds that can be parallelized.

The simplest examples of parallelizable manifolds are vector spaces and affine spaces. A very broad class of parallelizable manifolds is defined by any Lie group, and as Stiefel showed in his thesis, any compact, orientable 3-manifold is parallelizable. However, even when one gets into the world of homogeneous spaces, one does not have as many examples as one might expect. For instance, the only parallelizable spheres have dimensions 0, 1, 3, and 7, and similarly for projective spaces. In the three-dimensional case, a three-sphere $S^3$ is diffeomorphic to the manifold that underlies the Lie group $SU(2)$, while the three-dimensional projective space $\mathbb{R}P^3$ is diffeomorphic to the manifold of the Lie group $SO(3)$, and the projection $S^3 \to \mathbb{R}P^3$ that takes pairs of diametrically-opposite points $\{x, -x\}$ on the three-sphere to the line that goes through them is also the projection of $SU(2) \to SO(3)$ that one has to deal with in the quantum theory of angular momentum.

---

([1]) When no risk of ambiguity exists, we shall drop the explicit reference to *M*, for brevity.

At any rate, parallelizable manifolds are certainly adequate for the purposes of special relativistic models, although in the context of general relativity, one might potentially have to consider extending to "singular teleparallelism" to include the role of topological singularities of the space-time manifold.

*b. The geometry of parallelism.* – The relationship between a global frame field and the parallelism of lines in space is that when one has a global frame field $\mathbf{e}_i$ on a manifold $M$, one can say that if $x$ and $y$ are any two points in $M$ then a line $[\mathbf{v}]_x$ in the tangent space $T_x$ at $x$ is *parallel* to a line $[\mathbf{w}]_y$ in the tangent space $T_y$ at $y$ iff the components $v^i$ of any vector $\mathbf{v}$ that generates $[\mathbf{v}]_x$ with respect to the frame $\mathbf{e}_i(x)$ are proportional to the components $w^i$ of any vector $\mathbf{w}$ that generates $[\mathbf{w}]_y$ with respect to the frame $\mathbf{e}_i(y)$:

$$w^i = \lambda v^i, \qquad \text{for some real scalar } \lambda \neq 0. \tag{3.1}$$

When the proportionality constant $\lambda = 1$, one can say that the two vectors $\mathbf{v}$ and $\mathbf{w}$ are *equipollent*, although often they are still referred to as just "parallel." One can also express the condition (3.1) in terms of the reciprocal coframe field $\theta^i$ to $\mathbf{e}_i$:

$$\theta^i_x(\mathbf{v}) = \lambda\, \theta^i_y(\mathbf{w}). \tag{3.2}$$

From now on, we shall use only the special case of $\lambda = 1$, and say that tangent vectors are parallel when they have the same components in the two frames, and not just proportional ones.

The existence of a basis for each vector space $T_x$ and $T_y$ also means that for any $x, y \in M$, one has a linear isomorphism $P_{(x, y)} : T_x \cong T_y$, that is defined by making the basis $\mathbf{e}_i(x)$ correspond to the basis $\mathbf{e}_i(y)$. Hence, if $\mathbf{v} = \mathbf{e}_i(x) v^i$ is a tangent vector in $T_x$ then the tangent vector in $T_y$ that is parallel to $\mathbf{v}$ will be $\mathbf{e}_i(y) v^i$.

One can make analogous definitions for the parallelism of covectors.

One can define the map $P : M \times M \to T(M) \otimes T^*(M)$, $(x, y) \mapsto P_{(x, y)} \in T_y \otimes T_x^*$, which represents *parallel translation* under $\mathbf{e}_i$; in particular, it is defined globally for all possible pairs of points. Since $\theta^i_x$ is a basis for $T_x^*$ at each $x$ and $\mathbf{e}_i(y)$ is a basis for $T_y$ for each $y$, $\mathbf{e}_j(y) \otimes \theta^i_x$ will be a basis for $T_y \otimes T_x^*$ for each $(x, y)$. The map $P_{(x, y)}$ can be represented by the tensor field $\mathbf{e}_i(y) \otimes \theta_x^i$, since if $\mathbf{v} \in T_x$ then $\mathbf{e}_i(y) \, \theta_x^i(\mathbf{v}) = \mathbf{e}_i(y) \, v^i$, and the components of the linear isomorphism $P_{(x, y)}$ with respect to $\mathbf{e}_j(y) \otimes \theta_x^i$ are $\delta_j^i$ (i.e., the identity matrix) for every $(x, y)$. Of course, the transformation itself is not an identity map, since it maps between distinct vector spaces.

Note that this definition does not depend upon the action of any group of transformations on $M$, but when there is such a thing, one can often define parallel translation by differentiating that group action. For instance, in the cases of affine spaces and Lie groups, one has a simply transitive Lie group action on the space. In the former case, it is the action of the translation group on the points of the affine space, and in the latter, it is the action of the Lie group $G$ on itself $G \times G \to G$, $(g, g') \mapsto gg'$ by either left or right translation. That also has the effect of defining two distinct global frame fields on any $G$ that have the same frame at the identity element $e$, depending upon whether one left-translates the frame at the identity to every other point or right-translates it.

One can almost think of parallelizable manifolds as something like "almost Lie groups," when one generalizes the left or right translation that comes from the group structure to parallel translation when one no longer has a group structure in general. We shall see more indications to that effect below.

Having a global definition of parallelism and parallel translation allows one to define parallel vector fields and covector fields, as well as frame fields and coframe fields that are parallel to the one that is chosen initially. A vector field $\mathbf{v}(x) = \mathbf{e}_i(x) \, v^i(x)$ on $M$ is *parallel* (with respect to $\mathbf{e}_i$) iff its components $v^i(x)$ with respect to $\mathbf{e}_i$ are constant functions, or in other words:

$$\theta^i(\mathbf{v}) = v^i = \text{const.} \tag{3.3}$$

Dually, a coframe field $\alpha_x = \alpha_i(x) \, \theta_x^i$ is *parallel* (with respect to $\mathbf{e}_i$) iff its component functions $\alpha_i(x)$ are constants, which one can write as:

$$\alpha(\mathbf{e}_i) = \alpha_i = \text{const.} \tag{3.4}$$

The definition of frame fields and coframe fields that are parallel (with respect to $\mathbf{e}_i$) is simply the extension of the previous definitions to each of their member vector and covector fields. In particular, if $\bar{\mathbf{e}}_i$ is another global frame field on $M$ then one can express it in terms of $\mathbf{e}_i$ by way of the *transition function* $h : M \to GL(n)$, $x \mapsto h_j^i(x)$ as:

$$\bar{\mathbf{e}}_i(x) = \mathbf{e}_j(x) \, h_i^j(x) \quad \text{or simply} \quad \bar{\mathbf{e}}_i = \mathbf{e}_j \, h_i^j. \tag{3.5}$$

$\bar{\mathbf{e}}_i$ is *parallel to* $\mathbf{e}_i$ iff the transition function $h$ is a constant invertible matrix. One can also write this in terms of the reciprocal coframe field:

$$\theta^i(\bar{\mathbf{e}}_i) = h_i^j = \text{const.} \tag{3.6}$$

Hence, one can define an equivalence class [$\mathbf{e}_i$] of global frame fields under the parallelism that is defined by $\mathbf{e}_i$, and in fact, any other frame field of the equivalence class [$\mathbf{e}_i$] can serve as the definition of the same kind of parallelism.

Analogous statements pertain to a choice of coframe field to serve as the definition of parallelism, except that if that coframe field is reciprocal to $\mathbf{e}_i$ then one must invert the transition matrices in order to preserve contragredience.

One can also specialize the definition of parallel translation to parallel translation along a curve $x(t)$ in $M$: a vector field $\mathbf{v}(t)$ (covector field, frame field, coframe field) along $x(t)$ is *parallel-translated along* $x(t)$ iff its components with respect to $\mathbf{e}_i$ are constant:

$$\theta^i(\mathbf{v}(t)) = v^i(t) = \text{const.} \tag{3.7}$$

As a result, the derivatives of those components will also vanish:

$$\frac{dv^i}{dt} = 0. \tag{3.8}$$

In particular, if that is true of the velocity vector field of $x(t)$ then one can call that curve a *geodesic* (or perhaps "autoparallel" would be more precise). Hence, if the global frame field is one's reference frame for geometry then such a curve would appear "straight" with respect to the frame field. We shall shortly show how it makes a difference whether the frame field is "holonomic" or "anholonomic."

One can go to higher dimensions and define *geodesic submanifolds* of $M$ to be parallelizable submanifolds whose global frame fields are parallel with respect to the frame field on the ambient manifold $M$. Note that the example of a 2-sphere that is embedded in $\mathbb{R}^3$ shows that not all submanifolds of parallelizable manifolds have to be parallelizable.

*c. Other geometric constructions on parallelizable manifolds.* – Although parallelism is fundamentally an affine-geometric concept (which can be extended to projective geometry), nonetheless, the existence of a global frame field on a manifold $M$ allows one to define other geometric constructions as a consequence.

For instance, one can always define a volume element by way of the $n$-vector field:

$$\mathbf{V} = \mathbf{e}_1 \wedge \ldots \wedge \mathbf{e}_n = \frac{1}{n!} \varepsilon^{i_1 \cdots i_n} \mathbf{e}_{i_1} \wedge \cdots \wedge \mathbf{e}_{i_n} \tag{3.9}$$

or the $n$-form:

$$V = \theta^1 \wedge \ldots \wedge \theta^n = \frac{1}{n!} \varepsilon_{i_1 \cdots i_n} \theta^{i_1} \wedge \cdots \wedge \theta^{i_n}. \tag{3.10}$$

In these expressions the $\varepsilon$ symbol is the usual completely-antisymmetric Levi-Civita symbol that is $+1$, $-1$, or 0 depending upon whether the indices $i_1 \ldots i_n$ are an even or odd permutation of $12 \ldots n$, or not a permutation, respectively. The fact that $\mathbf{V}$ and $V$ are globally non-zero follows from the fact that the frame field is globally linearly-independent.

The definition of parallelism that relates to vector fields and covector fields extends to higher-order tensor fields, such as exterior forms: Such a tensor field is *parallel* with respect to the global frame field $\mathbf{e}_i$ iff its components with respect to the basis for the higher-order tensor product space in which the tensor takes its values are constant with respect to the basis. In particular, since $\mathbf{e}_1 \wedge \ldots \wedge \mathbf{e}_n$ defines a basis for the one-dimensional vector space $(\Lambda_n)_x$ of $n$-vectors, and the component of **V** is constant (viz., 1), one sees that the volume element **V** is parallel with respect to the chosen frame field. Analogous statements apply to *V*.

An obvious consequence of the foregoing definitions is that any parallelizable manifold must be orientable.

Of particular interest to relativity theory (or even Euclidian geometry, as one might find in Newtonian mechanics) is the fact that a global frame field always defines a metric tensor field of every signature type one might desire. Basically, one demands that the global frame field must be orthonormal for the resulting metric. If one defines the signature type of the metric by the $n \times n$ matrix $\eta_{ij} = \text{diag} (-1, \ldots, -1, +1, \ldots, +1)$, in which there are $p$ negative signs and $q$ positive ones, then one can define the metric by:

$$g = \eta_{ij}\, \theta^i\, \theta^j . \tag{3.11}$$

(The multiplication of covector fields is a symmetrized tensor product in this.)

One checks that:

$$g(\mathbf{e}_i, \mathbf{e}_j) = \eta_{ij}, \tag{3.12}$$

so the frame field $\mathbf{e}_i$ is indeed orthonormal for the metric thus-defined.

Since the symmetric tensor products $\theta^i\, \theta^j$ define a basis for the vector space $S_x^2$ of symmetric, second-rank covariant tensors at $x$, and the components of $g$ with respect to that basis are constants (viz., $\eta_{ij}$), one can say that the metric that is defined by (3.11) is parallel with respect to $\mathbf{e}_i$.

**4. Anholonomic frame fields.** – Any differentiable manifold $M$ is associated with an infinite-dimensional Lie algebra by way of its vector fields, at least when one regards a tangent vector as synonymous with the directional derivative operator that it defines, which is an operator that acts upon smooth functions on $M$. Namely, if $\mathbf{v}(x)$ is a tangent vector field on $M$ at $x$ and $f(x)$ is a smooth function on $M$ then the action of **v** on $f$ is defined by:

$$(\mathbf{v}f)(x) = df\,|_x(\mathbf{v}). \tag{4.1}$$

When this is expressed locally with respect to a natural frame for some coordinate chart $(U, x^i)$ on $M$, one will have:

$$(\mathbf{v}f)(x) = v^i(x) \frac{\partial f}{\partial x^i}\bigg|_x . \tag{4.2}$$

Although one can iterate the action of the first-order linear partial differential operator **v** and combine it with the action of other vector fields **w**, nonetheless, it is only the antisymmetric part of the product – i.e., the Lie bracket – that is useful in differential geometry:

$$[\mathbf{v}, \mathbf{w}]f = \mathbf{v}(\mathbf{w}f) - \mathbf{w}(\mathbf{v}f) = (v^j \partial_j w^i - w^j \partial_j v^i)\partial_i f. \tag{4.3}$$

That is, the components of the vector field $[\mathbf{v}, \mathbf{w}]$ with respect to the natural frame $\partial_i$ are:

$$[\mathbf{v}, \mathbf{w}]^i = v^j \partial_j w^i - w^j \partial_j v^i. \tag{4.4}$$

When $M$ is parallelizable and $\mathbf{e}_i$ is a global frame field on $M$, one can repeat this calculation with respect to $\mathbf{e}_i$, instead of a natural frame field and get:

$$[\mathbf{v}, \mathbf{w}]^i = v^i \mathbf{e}_i w^j - w^i \mathbf{e}_i v^j + v^i w^j [\mathbf{e}_i, \mathbf{e}_j]. \tag{4.5}$$

The first two terms look formally the same as before, but now there is an extra term that involves the Lie bracket $[\mathbf{e}_i, \mathbf{e}_j]$. In the case of a natural frame field, it would vanish, due to the equality of mixed second partial derivatives when one is dealing with smooth functions:

$$[\partial_i, \partial_j] = \frac{\partial^2}{\partial x^i \partial x^j} - \frac{\partial^2}{\partial x^j \partial x^i} = 0. \tag{4.6}$$

One would then says that a natural frame field is *holonomic*. More generally, a frame field $\mathbf{e}_i$ is holonomic iff:

$$[\mathbf{e}_i, \mathbf{e}_j] = 0. \tag{4.7}$$

The question then arises of whether holonomic frame fields always represent natural frame fields for some coordinate system. It turns out that this is true locally, but an elementary counterexample to the extension of that to a global statement is any compact, Abelian group, such as a torus. Although any left (or right) invariant frame field on such a thing will be holonomic, no compact manifold can be diffeomorphic to $\mathbb{R}^n$ for any $n > 0$.

However, not all frame fields are holonomic. More generally, there will always be functions $c_{ij}^k(x)$ on $M$ such that:

$$[\mathbf{e}_i, \mathbf{e}_j] = c_{ij}^k \mathbf{e}_k, \tag{4.8}$$

since any vector field on $M$, and in particular, $[\mathbf{e}_i, \mathbf{e}_j]$, can be expressed in terms of its components with respect to $\mathbf{e}_k$. When the $c_{ij}^k$ do not all vanish, the frame field $\mathbf{e}_i$ is then said to be *anholonomic.*

Due to the antisymmetry of the Lie bracket, one will also have the antisymmetry of $c_{ij}^k$ in the lower indices:

$$c_{ij}^k = -c_{ji}^k. \tag{4.9}$$

The functions $c_{ij}^k(x)$ are called the *structure functions* of the frame field $\mathbf{e}_i$. If they bear a striking resemblance to the structure *constants* of any Lie group then that is because defining a global frame field on a Lie group by left or right translation has the effect of making those functions into constant functions; i.e., the structure constants of a Lie group are the structure functions of a left (or right) invariant global frame field on it.

Once again, one might consider the converse problem, namely, is every parallelizable manifold that admits a global frame field with constant structure functions diffeomorphic to some Lie group? Apparently, this is true locally (cf., [20], where it is stated without proof). Hence, we have further support for our suspicion that parallelizable manifolds are, in some sense, "almost Lie groups."

The Jacobi identity for the Lie algebra of vector fields on *M*, namely:

$$0 = [\mathbf{u}, [\mathbf{v}, \mathbf{w}]] + [\mathbf{v}, [\mathbf{w}, \mathbf{u}]] + [\mathbf{w}, [\mathbf{u}, \mathbf{v}]], \tag{4.10}$$

will introduce some extra terms in the case of an anholonomic frame field, due to the fact that if *f* is a smooth function on *M* then:

$$[\mathbf{v}, f\mathbf{w}] = f[\mathbf{v}, \mathbf{w}] + (\mathbf{v}f)\,\mathbf{w}. \tag{4.11}$$

When this is applied to $[\mathbf{e}_i, [\mathbf{e}_j, \mathbf{e}_k]]$, one will get:

$$[\mathbf{e}_i, [\mathbf{e}_j, \mathbf{e}_k]] = [\mathbf{e}_i, c_{jk}^l \mathbf{e}_l] = c_{jk}^l [\mathbf{e}_i, \mathbf{e}_l] + (\mathbf{e}_i c_{jk}^l)\,\mathbf{e}_l = (c_{jk}^m c_{im}^l + \mathbf{e}_i c_{jk}^l)\,\mathbf{e}_l.$$

Upon cyclic permutation of *ijk*, one gets the *generalized Jacobi identity:*

$$0 = c_{jk}^m c_{im}^l + c_{ki}^m c_{jm}^l + c_{ij}^m c_{km}^l + \mathbf{e}_i c_{jk}^l + \mathbf{e}_j c_{ki}^l + \mathbf{e}_k c_{ij}^l. \tag{4.12}$$

One sees that when the structure functions are constants, such as on a Lie group, the last three terms will disappear, and one will be left with the usual statement for the Jacobi identity in terms of the structure constants.

If $\theta^i$ is the reciprocal coframe field to $\mathbf{e}_i$ then the statements above can be given dual formulations in terms of $\theta^i$. First, one uses the "intrinsic" definition for the exterior derivative of a 1-form $\alpha$, namely, if **v** and **w** are vector fields then ([1]):

$$d_\wedge \alpha(\mathbf{v}, \mathbf{w}) = \mathbf{v}(\alpha(\mathbf{w})) - \mathbf{w}(\alpha(\mathbf{v})) - \alpha([\mathbf{v}, \mathbf{w}]). \tag{4.13}$$

One then applies this to $\theta^i$ while using the vector fields of $\mathbf{e}_i$:

$$\begin{aligned}d_\wedge \theta^i(\mathbf{e}_j, \mathbf{e}_k) &= \mathbf{e}_j(\theta^i(\mathbf{e}_k)) - \mathbf{e}_k(\theta^i(\mathbf{e}_j)) - \theta^i([\mathbf{e}_j, \mathbf{e}_k]) \\ &= \mathbf{e}_j \delta_k^i - \mathbf{e}_k \delta_j^i - c_{jk}^l \theta^i(\mathbf{e}_l)\end{aligned}$$

---

([1]) We prefer to use a separate notation for the ordinary differential *d* of a differentiable map and the exterior derivative $d_\wedge$ of a differential form, since we will have to use both in this study.

$$= - c^i_{jk},$$

since the Kronecker delta's are constants. However, one also has that $d_\wedge \theta^i (\mathbf{e}_j, \mathbf{e}_k)$ are the components of $d_\wedge \theta^i$ with respect to the coframe $\theta^i$, so:

$$d_\wedge \theta^i = -\tfrac{1}{2} c^i_{jk} \theta^j \wedge \theta^k. \tag{4.14}$$

Once again, this is a generalization of the *Maurer-Cartan* equations for a Lie group, for which the $c^i_{jk}$ would be constant. These equations are also dual to the defining equations (4.8) of the structure functions.

One can also find the dual to the generalized Jacobi identity by taking the exterior derivative of both sides of (4.14), while keeping in mind that the $c^i_{jk}$ are not necessarily constants:

$$0 = -\tfrac{1}{2} dc^i_{jk} \wedge \theta^j \wedge \theta^k - \tfrac{1}{2} c^i_{jk} d_\wedge \theta^j \wedge \theta^k + \tfrac{1}{2} c^i_{jk} \theta^j \wedge d_\wedge \theta^k.$$

Using the fact that:

$$dc^i_{jk} = \partial_l c^i_{jk} dx^l = h_l^m \partial_m c^i_{jk} \theta^l = (\mathbf{e}_l c^i_{jk}) \theta^l$$

and the generalized Maurer-Cartan equations (4.14), one gets:

$$0 = -\tfrac{1}{3} [\mathbf{e}_i c^l_{jk} + \mathbf{e}_j c^l_{ki} + \mathbf{e}_k c^l_{ij} + c^m_{ij} c^l_{km} + c^m_{jk} c^l_{im} + c^m_{ki} c^l_{jm}] \theta^i \wedge \theta^j \wedge \theta^k, \tag{4.15}$$

which is essentially the same thing as the generalized Jacobi identity above.

Note that saying that:

$$df = (\mathbf{e}_i f) \theta^i \tag{4.16}$$

is more general that saying that $df = \partial_i f\, dx^i$, since there are parallelizable manifolds that do not admit global coordinate systems for global natural frame fields to exist.

**5. Teleparallelism.** – We shall now move on to the traditional way of discussing the geometry of parallelizable manifolds, which involves introducing a connection that will make the global frame field $\mathbf{e}_i$ parallel. Since the traditional discussions of teleparallelism in relativity tend to always assume that there is also a natural frame field for a coordinate chart, there will then be two ways of describing the components of the connection, along with its torsion and curvature, depending upon whether one is using the holonomic (i.e., natural) frame field or the anholonomic one as a reference.

*a. The teleparallelism connection.* – The real issue is what one must do with the expressions $d\mathbf{e}_i$ and $d\theta^i$, where the $d$ refers to the differential, not the exterior derivative. If $\mathbf{e}_i = \partial_j \tilde{h}_i^j$ then differentiation will give:

$$d\mathbf{e}_i = \partial_j \otimes d\tilde{h}_i^j = \mathbf{e}_j \otimes h_k^j \, d\tilde{h}_i^k = \mathbf{e}_j \otimes \omega_i^j, \tag{5.1}$$

in which we have introduced the *teleparallelism connection 1-form:*

$$\omega_j^i = h_k^i \, d\tilde{h}_j^k = -dh_k^i \, \tilde{h}_j^k. \tag{5.2}$$

(The second equality is obtained by differentiating the equation $h_k^i \tilde{h}_j^k = \delta_j^i$.)

Note that this type of connection is a generalization of the usual way that one introduces the angular velocity of an orthonormal frame that rotates with respect to another one in time, in that the time derivative has been replaced with partial derivatives, in effect. It is also a generalization of the matrix that enters into the Frenet-Serret equations for a moving orthonormal frame along a curve to higher-dimensional objects than curves. (See the author's discussion of that idea in [**21**].)

By introducing the *covariant differential* operator, one can also write (5.1) in the form:

$$\nabla \mathbf{e}_i \equiv d\mathbf{e}_i - \mathbf{e}_j \otimes \omega_i^j = 0, \tag{5.3}$$

which shows that $\mathbf{e}_i$ is indeed parallel for that connection.

In order to account for the fact that the components of a parallel vector field $\mathbf{v}$ with respect to $\mathbf{e}_i$ must be constant, let us represent $\mathbf{v}$ in both the holonomic and anholonomic frames $\mathbf{v} = \partial_i v^i = \mathbf{e}_i \bar{v}^i$, with $\bar{v}^i = h_j^i v^j$. Differentiation the first equality gives:

$$d\mathbf{v} = \partial_i \otimes dv^i = \mathbf{e}_i \otimes h_j^i \, dv^j.$$

Differentiating the second equality and using (5.1) gives:

$$d\mathbf{v} = \mathbf{e}_i \otimes (d\bar{v}^i + \omega_j^i \bar{v}^j) = \mathbf{e}_i \otimes (d\bar{v}^i - h_k^i \omega_j^k v^j).$$

Equating the two expressions (which must be true for all $\mathbf{v}$) will give:

$$d\bar{v}^i = h_j^i (dv^j + \omega_k^j v^k) = h_j^i \nabla v^j, \tag{5.4}$$

in which we have defined the covariant differential of $v^i$ by:

$$\nabla v^i \equiv dv^j + \omega_k^j v^k. \tag{5.5}$$

Hence, $d\bar{v}^i$ will vanish iff $\nabla v^i$ vanishes.

Since connection 1-forms do not transform like tensors, one must distinguish the form that $\omega_j^i$ takes in the holonomic frame field from the form $\bar{\omega}_j^i$ that it takes in the anholonomic one. In fact the transformation becomes:

$$\bar{\omega}^i_j = \tilde{h}^i_k \omega^k_l h^l_j + \tilde{h}^i_k \, dh^k_j = \tilde{h}^i_k (h^k_l d\tilde{h}^l_k) h^k_j + \tilde{h}^i_k \, dh^k_j = d\tilde{h}^i_k h^k_j + \tilde{h}^i_k \, dh^k_j = d(\tilde{h}^i_k h^k_j),$$

so

$$\bar{\omega}^i_j = 0. \tag{5.6}$$

Hence, if we define the covariant differential of the components $\bar{v}^i$ by:

$$\nabla \bar{v}^i \equiv d\bar{v}^i + \bar{\omega}^i_j \bar{v}^j = d\bar{v}^i \tag{5.7}$$

then we will see that (5.4) can be given the form:

$$\nabla \bar{v}^i = h^i_j \nabla v^j, \tag{5.8}$$

which justifies the term "covariant" derivative.

Since the vanishing of the connection 1-form is typical of Euclidian and Minkowskian geometries, one sees that the geometry of teleparallelism is simply elementary geometry when one refers everything to an anholonomic frame, instead of a holonomic one.

A dual argument that addresses the reciprocal coframe $\theta^i$ and the components of a 1-form $\alpha = \alpha_i \, dx^i = \bar{\alpha}_i \theta^i$ gives dual results to the ones above:

$$\nabla \theta^i \equiv d\theta^i + \omega^i_j \theta^j = 0, \tag{5.9}$$

which shows that the reciprocal coframe field to a parallel frame field is parallel, and:

$$\nabla \alpha_i \equiv d\alpha_i - \alpha_j \omega^j_i = h^i_j \, d\bar{\alpha}^j, \tag{5.10}$$

so one can also say that a 1-form is parallel iff its coefficients are constant in the anholonomic frame or its covariant differential vanishes in the holonomic one.

*b. Geodesics.* – In particular if **v** (*t*) represents the velocity vector field of a curve *x* (*t*) in *M* then the geodesic equation will take either the form:

$$\nabla_{\mathbf{v}} v^i = \frac{dv^i}{dt} + \omega^i_j(\mathbf{v}) v^j = 0, \tag{5.11}$$

which pertains to the holonomic frame field, or:

$$\nabla_{\mathbf{v}} \bar{v}^i = \frac{d\bar{v}^i}{dt} = 0, \tag{5.12}$$

which pertains to the anholonomic one.

If we introduce the components $\omega^i_{kj}$ of $\omega^i_j$ with respect to $dx^i$:

$$\omega^i_j = \omega^i_{kj} \, dx^k \tag{5.13}$$

then (5.11) can be given the conventional component form:

$$\frac{dv^i}{dt} + \omega^i_{jk} v^j v^k = 0. \tag{5.14}$$

*c. The Cartan structure equations for teleparallelism.* – The Cartan structure equations for the teleparallelism connection are straightforward to calculate with respect to both frame fields. In the holonomic case, one has the structure equation for torsion in the form:

$$\Theta^i = d_\wedge(dx^i) + \omega^i_j \wedge dx^j = \tfrac{1}{2}(\omega^i_{jk} - \omega^i_{kj}) \, dx^j \wedge dx^k, \tag{5.15}$$

so if one also expresses $\Theta^i$ in terms of its components with respect to $dx^i$ as:

$$\Theta^i = \tfrac{1}{2} S^i_{jk} \, dx^j \wedge dx^k \tag{5.16}$$

(in which the choice of symbol is based in convention) then one can say that:

$$S^i_{jk} = \omega^i_{jk} - \omega^i_{kj} ; \tag{5.17}$$

i.e., in a holonomic frame, the torsion of the connection is simply due to the antisymmetric part of its components.

The structure equation for curvature in the holonomic frame field takes the form:

$$\begin{aligned}\Omega^i_j &= d_\wedge \omega^i_j + \omega^i_k \wedge \omega^k_j \\ &= d_\wedge (h \, d\tilde{h})^i_j + (h \, d\tilde{h})^i_k \wedge (h \, d\tilde{h})^k_j = (dh \wedge d\tilde{h})^i_j + (h \, d\tilde{h} \, h \wedge d\tilde{h})^i_j \\ &= (dh \wedge d\tilde{h})^i_j - (dh \wedge d\tilde{h})^i_j,\end{aligned}$$

which gives:

$$\Omega^i_j = 0. \tag{5.18}$$

This was to be expected, since the vanishing of curvature is necessary for the existence of parallel vector fields.

In the anholonomic frame, one has the structure equation for torsion in the form:

$$\Theta^i = d_\wedge \theta^i + \bar{\omega}^i_j \wedge \theta^j = d_\wedge \theta^i, \tag{5.19}$$

so if the components of $\Theta^i$ in the anholonomic form are defined by:

$$\Theta^i = \tfrac{1}{2} \bar{S}^i_{jk} \, \theta^j \wedge \theta^k \tag{5.20}$$

then one can say that:

$$\overline{S}^i_{jk} = -c^i_{jk} \; ; \tag{5.21}$$

i.e., the components of the torsion 2-form with respect to the anholonomic frame field are minus the structure functions.

The structure equation for curvature becomes trivial in the anholonomic frame:

$$\Omega^i_j = d_\wedge \overline{\omega}^i_j + \overline{\omega}^i_k \wedge \overline{\omega}^k_j = 0 . \tag{5.22}$$

*d. The Bianchi identities for teleparallelism.* – We have seen before that the generalized Bianchi identities for the structure functions of an anholonomic coframe field $\theta^i$ are dual to the generalized Jacobi identities for the structure functions, when they are defined by the reciprocal frame field $\mathbf{e}_i$, and that they can be obtained by simply exterior differentiating the defining equations for the structure functions of $\theta^i$.

In the present case, the only thing that changes is that the usual definition of the Bianchi identities involves the exterior covariant derivative, instead of the exterior derivative. When one applies it to the Cartan structure equations in their general sense, one gets:

$$\nabla_\wedge \Theta^i = d_\wedge \Theta^i + \omega^i_j \wedge \Theta^j = d_\wedge(d_\wedge \theta^i + \omega^i_j \wedge \theta^j) + \omega^i_j \wedge (d_\wedge \theta^j + \omega^j_k \wedge \theta^k) ,$$

which simplifies to:

$$\nabla_\wedge \Theta^i = \Omega^i_j \wedge \theta^j. \tag{5.23}$$

When the curvature vanishes, this becomes:

$$\nabla_\wedge \Theta^i = 0 \tag{5.24}$$

or

$$d_\wedge \Theta^i = -\omega^i_j \wedge \Theta^j . \tag{5.25}$$

In the holonomic frame field, for which one has $\Theta^i = \omega^i_j \wedge dx^j$, (5.24) takes the form:

$$(d_\wedge \omega^i_j + \omega^i_k \wedge \omega^k_j) \wedge dx^j = \Omega^i_j \wedge dx^j = 0, \tag{5.26}$$

from the vanishing of curvature.

In the anholonomic frame field, $\Theta^i = d_\wedge \theta^i$ and $\omega^i_j = \overline{\omega}^i_j = 0$, so the Bianchi identity for torsion is once again derived from the identity that comes from taking the exterior derivative of torsion, rather than the exterior covariant derivative.

The general form of the Bianchi identity for curvature is:

$$\nabla_\wedge \Omega^i_j = d_\wedge \Omega^i_j + \omega^i_k \wedge \Omega^k_j = \Omega^i_k \wedge \omega^k_j . \tag{5.27}$$

Of course, since the curvature 2-form vanishes with respect to either frame field in the present case of teleparallelism, that will be trivial.

*e. Covariant differentials of the volume element and metric.* – As we pointed out above, the volume element *V* and metric *g* (of any signature type) that are defined by the coframe field $\theta^i$ are parallel with respect to that coframe field; i.e., their components with respect to it (namely, 1 and $\eta_{ij}$, resp.) are constants. We shall now show that their covariant differentials vanish accordingly.

First, we express the volume element in terms of both the anholonomic and holonomic frame fields:

$$V = \theta^1 \wedge \ldots \wedge \theta^n = (\det h) \, dx^1 \wedge \ldots \wedge dx^n. \tag{5.28}$$

We then take the differential of *V* and express it in both ways. First, in terms of $\theta^i$:

$$dV = d\theta^1 \wedge \ldots \wedge \theta^n + \ldots + \theta^1 \wedge \ldots \wedge d\theta^n$$
$$= (\omega^1_i \otimes \theta^i) \wedge \ldots \wedge \theta^n + \theta^1 \wedge \ldots \wedge (\omega^n_i \otimes \theta^i).$$

Since the only $\theta^i$ that will produce a non-zero *n*-form in each terms of this sum is the one that corresponds to the $\theta^i$ that is being differentiated, the latter expression will become:

$$dV = (\mathrm{Tr}\, \omega) \otimes V. \tag{5.29}$$

Of course, the representation of $\omega$ that we must use in the anholonomic frame field is $\bar{\omega}$, which vanishes, so its trace will, as well. Hence:

$$dV = 0. \tag{5.30}$$

We now take expression *dV* in terms of the holonomic frame field:

$$dV = d(\det h) \otimes dx^1 \wedge \ldots \wedge dx^n. \tag{5.31}$$

As one can verify:

$$d(\det h) = (\mathrm{Tr}\, \omega) \det h, \tag{5.32}$$

and since *h* is always invertible, the right-hand side of this will vanish iff $\mathrm{Tr}\, \omega$ vanishes again.

We can then say that the connection 1-form $\omega$ takes its values in Lie algebra $\mathfrak{sl}(n)$.

Now, let us look at the covariant differential of the metric. First we express it in terms of both the anholonomic and holonomic frame fields:

$$g = \eta_{ij} \, \theta^i \theta^j = g_{ij} \, dx^i \, dx^j. \tag{5.33}$$

In the first case:

$$dg = d\eta_{ij} \, \theta^i \theta^j + \eta_{ij} \, d\theta^i \, \theta^j + \eta_{ij} \, \theta^i \, d\theta^j = (\eta_{ik} \omega^k_j + \eta_{kj} \omega^k_i) \otimes \theta^i \theta^j.$$

Of course, once again the connection form $\omega$ will vanish in the anholonomic frame field, so:

$$dg = 0. \tag{5.34}$$

In the holonomic frame field, we get:

$$dg = dg_{ij} \otimes dx^i \, dx^j - g_{ij} \, (\omega^i_k \otimes dx^k) \, dx^j - g_{ij} \, dx^i \, (\omega^j_k \otimes dx^k) = \nabla g_{ij} \otimes dx^i \, dx^j,$$

in which we have introduced the *covariant differential* of $dg_{ij}$:

$$\nabla g_{ij} = dg_{ij} - g_{ik} \, \omega^k_j - g_{kj} \, \omega^k_i . \tag{5.35}$$

Since $dg$ vanishes, so must this expression, which is consistent with the previous statement that $g$ was parallel under $\theta^i$.

**6. Discussion.** – Since the main purpose of this article was to serve as a primer on the geometry of parallelizable manifolds, as it is described in Cartan's language of differential forms and moving frames, and especially to relate that geometry to the properties of anholonomic frame fields, an obvious direction to pursue would be to apply the methods used here to the problem of teleparallelism as a field theory of gravitation. Here, the approach of Itin, *et al.* [**14**], would seem to be the most promising, since it is an extension of the methods of Maxwell's equations of electromagnetism, which are also quite rooted in the calculus of differential forms.

Another developing application of teleparallelism is to continuum mechanics. This line of research mostly goes back to the work of Kondo, *et al.* [**22**], although the term "teleparallelism" was never used explicitly in that reference; the author also made some first observations along those lines in [**21**]. The role of dislocations as sources of internal stresses has been established for some time, and their roots in topology goes back to the seminal paper by Volterra [**23**] (see also the author's comments on Volterra's theorem [**24**]). The author pointed out in [**21**] that since the de Rham cohomology in dimension one relates to the existence of closed 1-forms that are not exact, when those 1-forms are the legs of a global coframe field, the resulting teleparallelism connection will not only have vanishing curvature, as usual, but also vanishing torsion; i.e., it will be *flat*. However, that does not have to imply a trivial holonomy group, but only a discrete one.

Another direction that one might wish to pursue is the geometry of anholonomic spaces [**25-27**], which are not only rooted in the theory of mechanics with non-holonomic constraints ([1]), but have also been proposed as another way of unifying Einstein with Maxwell. Basically, the issue relates to whether the space-time manifold is something that exists as an integral submanifold of a higher-dimensional space only in the case of an integrable frame field, which would imply the holonomic case, and that the anholonomic frame field would then become more fundamental than the points that it projects onto.

Of course, the possibility that was initially rejected for the sake of this study, namely, that the manifold upon which one wished to define a frame field was not parallelizable, must be addressed eventually. One would then have to work with "singular" frame fields; i.e., ones that are not defined globally, and especially ones that are defined on "maximal" subsets, such as the complement of a finite set of points. Although one cannot extend the partial frame field to a global one in such a case, one can always extend the partial connection to a global one, but not always canonically. (Typically, one uses a partition of unity.) If the two-sphere is any indication then perhaps one of the things that could change when one goes to "singular teleparallelism" is that the connection that is defined by a singular frame field might have non-zero curvature at the singularities. A particularly interesting class of non-parallelizable manifolds that includes the 2-sphere and the 4-sphere is defined by the "suspensions" of parallelizable manifolds. That process involves first extending a parallelizable manifold to a cylinder of finite length (which will also be parallelizable) and then identifying the boundary components of the cylinder to points, which will then become singular points of any frame field on the cylinder. Suspending the circle and the 3-sphere in that way will produce the 2-sphere and the 4-sphere, respectively.

---

([1]) Apparently, the main distinction that dictates whether one will use the adjective "anholonomic" or "non-holonomic" is whether one is dealing with the literature of relativity or mechanics, respectively.

### References ([1])

---

([1]) References marked with an asterisk are available in English translation at neo-classical-physics.info.

__________